# Millisecond-scale motor encoding in a cortical vocal area

Claire Tang[1,2], Diala Chehayeb[2], Kyle Srivastava[3], Ilya Nemenman[2,4], and Samuel Sober[2]

[1]Neuroscience Graduate Program, University of California, San Francisco, San Francisco, CA 94158, USA
[2]Department of Biology, Emory University, Atlanta, GA 30322, USA
[3]Department of Biomedical Engineering, Georgia Institute of Technology, Atlanta, GA 30332, USA
[4]Department of Physics, Emory University, Atlanta, GA 30322, USA

**Abstract:**

Studies of motor control have almost universally examined firing rates to investigate how the brain shapes behavior. In principle, however, neurons could encode information through the precise temporal patterning of their spike trains as well as (or instead of) through their firing rates. Although the importance of spike timing has been demonstrated in sensory systems, it is largely unknown whether timing differences in motor areas could affect behavior. We tested the hypothesis that significant information about trial-by-trial variations in behavior is represented by spike timing in the songbird vocal motor system. We found that premotor neurons convey information via spike timing far more often than via spike rate and that the amount of information conveyed at the millisecond timescale greatly exceeds the information available from spike counts. These results demonstrate that information can be represented by spike timing in motor circuits and suggest that timing variations evoke differences in behavior.

**Introduction:**

The relationship between patterns of neural activity and the behaviorally relevant parameters they encode is a fundamental problem in neuroscience. Broadly speaking, a neuron might encode information in its spike rate (the total number of action potentials produced) or in the fine temporal pattern of its spikes. In sensory systems as diverse as vision, audition, somatosensation, and taste, prior work has demonstrated that information about stimuli can be encoded by fine temporal patterns, in some cases where no information can be detected in a rate code [1-11]. This information present in fine temporal patterns might be decoded by downstream areas to produce meaningful differences in perception or behavior.

However, in contrast to the extensive work on temporal coding in sensory systems, the timescale of encoding in forebrain motor networks has not been explored. It is therefore unknown whether the precise temporal encoding observed in sensory systems is propagated to cortical motor circuits or whether millisecond-scale spike timing differences in motor networks could result in differences in behavior. Although many studies have shown that firing rates can predict variations in motor output [12-14], to our knowledge no studies have examined whether different spiking patterns in cortical neurons evoke different behavioral outputs even if the firing rate remains the same.

The songbird provides an excellent model system for testing the hypothesis that fine temporal patterns in cortical motor systems can encode behavioral output. Song acoustics are modulated on a broad range of time scales, including fast modulations on the order of 10 msec [15,16]. Vocal patterns are organized by premotor neurons in vocal motor cortex (the robust nucleus of the arcopallium, or RA; Fig. 1a), which directly synapse with motor neurons innervating the vocal muscles[14,15,17]. Bursts of action potentials in RA (Fig. 1b) are precisely locked in time to production of vocal gestures ("song syllables"), suggesting that the timing of bursts is tightly controlled [18]. Similarly, the ensemble activity of populations of RA neurons can be used to estimate the time during song with approximately 10 msec uncertainty [15]. However, although these prior studies demonstrate that the timing of bursts is tightly aligned to the timing of song syllables, it is unknown how the temporal patterns of spikes within bursts might encode the trial-by-trial modulations in syllable acoustics known to underlie vocal plasticity [19]. Significantly, biomechanical studies have shown that vocal muscles in birds initiate and complete their force production within a few milliseconds of activation (far faster than those seen in most mammalian skeletal muscles), suggesting that RA's downstream targets can transduce fine temporal spike patterns into meaningful differences in behavior [20,21]. However, while it is clear that trial-by-trial variation in spike counts can predict variations in the acoustics of individual song syllables [14,22], it is unknown whether the precise timing of spikes within bursts might be even better predictors of vocal motor output than spike counts.

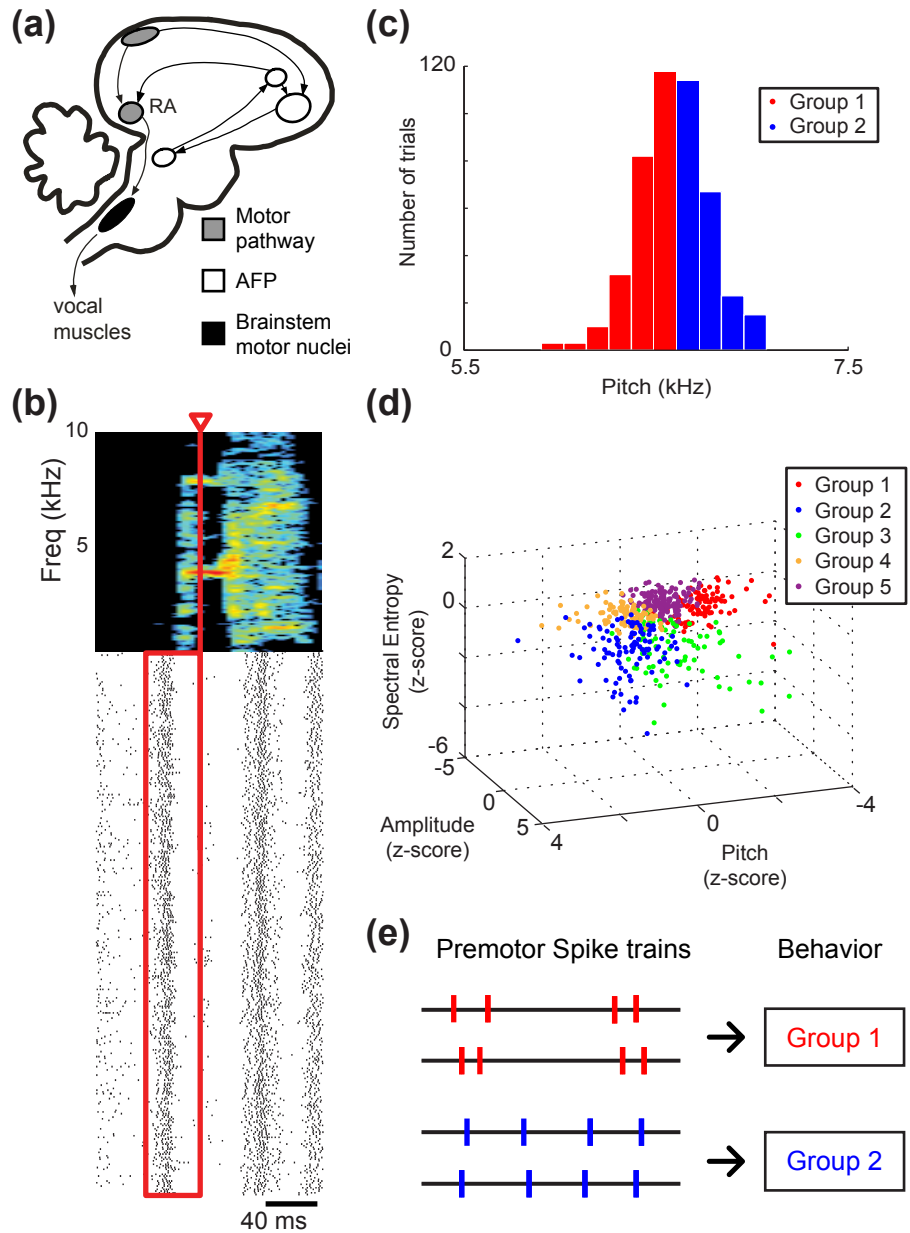

To quantify the temporal scale of encoding in the vocal motor system, we adapted well-established mathematical tools that have previously been applied to measure information transfer in sensory systems. First, we used a spike train distance metric to quantify the differences between pairs of spike trains produced during different renditions of individual song syllables and a classification scheme to quantify whether distance metrics based on rate or timing yielded the best prediction of acoustic output [23,24].

**Figure 1. Neural data and spike train analysis. a,** The song system consists of two pathways, the direct motor pathway and the anterior forebrain pathway (AFP). Neurons in premotor nucleus RA project to brainstem motor neurons that innervate the vocal muscles. **b,** Spike trains recorded from a single RA neuron. Spectrogram of a single song syllable at top shows the acoustic power (color scale) at different frequencies as a function of time. Each tick mark (bottom) represents one spike and each row represents one iteration of the syllable. We analyzed spikes produced in a 40 msec premotor window (red box) prior to the time when acoustic features were measured (red arrow). **c,** Syllable iterations divided into categories ("behavioral groups") based on a single acoustic parameter. Here, iterations of a song syllable were divided into two groups ($N$=2; see Methods) based on fundamental frequency ("pitch"). **d,** Syllable iterations divided into $N$=5 groups by k-means clustering in a three-dimensional acoustic parameter space **e,** We asked whether spike trains could be used to predict differences in behavior. Specifically, our analysis quantifies the extent to which differences in spike timing can discriminate the behavioral group from which the trial was drawn. This is shown in the schematic, in which differences in spike timing contain information about behavioral group even if spike counts (four spikes in this example) are identical across trials.

Second, we used model-independent information theoretic methods to compute the mutual information between spike trains and acoustic features of vocal behavior [8,10]. Crucially, both techniques measure information present in the neural activity at different timescales, allowing us to quantify the extent to which spike timing in motor cortex predicts upcoming behavior.

**Results:**

We collected extracellular recordings from projection neurons in vocal motor area RA in songbirds (Fig. 1a). In total, we analyzed 34 single-unit cases and 91 multiunit cases, where a "case" is defined as a neural recording being active prior to the production of a syllable (Fig. 1b), as explained in Methods. The number of trials (syllable iterations) recorded in each case varied from 51 to 1003 (median 115, mean 192.4). Iterations of each song syllable were divided into groups based on acoustic similarity ("behavioral groups"; Fig. 1c-d), and information-theoretic analyses were used to quantify whether the timing of spikes within bursts conveys significant information about upcoming motor output, as schematized in Figure 1e.

Metric-space analysis

We first used a version of the metric-space analysis established by Victor and Purpura to compare the information conveyed by spike rate and spike timing [24,25]. As described in Methods, this analysis quantifies how mutual information between neural activity and motor output depends on a cost parameter $q$, which quantifies the extent to which spike timing (as opposed to spike number) contributes to the dissimilarity, or "distance", between spike trains (Fig. 2a). The distance between two spike trains is computed by quantifying the cost of transforming one spike train into the other. Here, parameter $q$, measured in msec$^{-1}$, quantifies the relative cost of changing spike timing by 1 msec, as compared to the fixed cost of 1.0 for adding or subtracting a spike. Spike train distances are then used to classify iterations of each song syllable into behavioral groups, and the performance of the classifier $I(G^P, G)$ is used to quantify the mutual information between neural activity and vocal output. Figure 2b shows a representative "rate case", where $q_{max}=0$ (that is, information is maximized at $q = 0$, where spike train distances are computed based solely on spike counts). As $q$ increases, the performance of the classifier decreases from its maximal value. This means that the best discrimination between behavioral groups (Fig. 1c-d) occurs when only spike counts are used in calculating the distances between pairs of spike trains. In contrast, Figure 2c illustrates a "temporal case". In temporal cases, mutual information between neural activity and vocal motor output reaches its peak when $q > 0$. This indicates that there is better discrimination when spike timings are taken into consideration. Note that in the case shown in Figure 2c, the rate code does not provide significant information about behavioral output (empty symbol at $q=0$).

Across all analyses in cases where information was significant at any value of $q$, including cases where $q_{max} = 0$, the median value of $q_{max}$ was 0.3, suggesting a high prevalence of temporal cases. Figure 2d shows the prevalence of rate cases and temporal cases in our dataset. As described in Methods, we assigned the iterations of each song syllable to behavioral groups based either on a single acoustic parameter (e.g. pitch, Fig. 1c) or using multidimensional clustering ("3D acoustics", Fig. 1d). The different grouping techniques yielded similar results. When syllable acoustics were grouped by clustering in a three-dimensional parameter space (Fig. 2d, blue bars) the fraction of temporal cases was significantly greater than the fraction of rate cases (blue asterisk; $p<10^{-8}$, z-test for proportions). Similarly, temporal cases significantly outnumbered rate cases when acoustics were grouped using only a single parameter (pitch, amplitude, or spectral entropy, shown by green, yellow, and red asterisks respectively; $p<10^{-8}$). Note that in some cases these analyses did not yield a significant value of $I(G^P, G; q)$ for any value of $q$ and thus were neither rate nor temporal cases; therefore the fractions in Figure 2d do not sum to unity. Additionally, we asked whether the proportions of temporal cases shown in Figure 2d were significantly greater than chance by randomizing the spike times in each trial ("Poisson test"; Methods). This analysis revealed a significant proportion of temporal cases when vocal acoustics were measured by multidimensional clustering ("3D acoustics", $p<0.05$ after Bonferroni correction for multiple comparisons indicated by cross in Fig. 2d) but the same measure fell short of significance when the three acoustic parameters were considered individually ($p=0.06-0.24$ after Bonferroni correction).

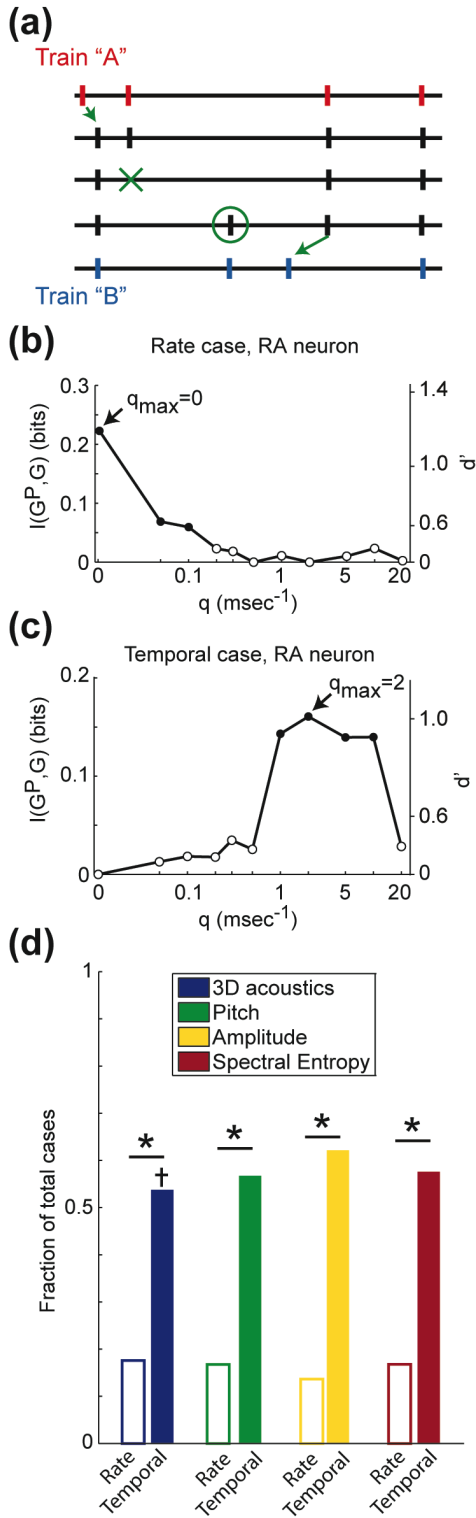

**Figure 2. Metric-space analysis reveals temporal coding in the vocal motor system. a,** The distance between example spike trains "A" and "B" is the sum of the fixed costs of adding and subtracting spikes (green circle and "X", respectively) and the cost of changing spike timing (green arrows), which is parameterized by the constant $q$ (see Methods). **b,** Representative rate case. Information $I(G^P,G)$ about upcoming vocal behavior is maximized when $q = 0$, indicating a rate code. Filled circles, information significantly greater than zero; empty circles, nonsignificant values. In this case, acoustically similar syllable renditions ("behavioral groups") were grouped by spectral entropy. **c,** Representative temporal case. Here information is maximized when $q > 0$, indicating a temporal code. Note that there is no information in the spike count (unfilled circle at $q = 0$). In this example, syllables were grouped by pitch. Right-hand vertical axes in **b** and **c** shows information values converted into d' units (note nonlinear scale). **d,** Prevalence of rate and temporal cases. For each acoustic grouping, the proportion of temporal cases is significantly greater than the proportion of rate cases (asterisks, $p<10^{-8}$, z-tests for proportions). Proportions of rate and temporal cases do not differ significantly across the four behavioral groupings. Furthermore, the proportion of temporal cases is significantly greater than that expected by chance for 3D acoustics (cross, $p<0.05$, Poisson test with Bonferroni correction). In all analyses shown, the maximum possible information is 1 bit ($N=2$ behavioral groups; see Methods), which corresponds to perfect discrimination between groups.

To measure the maximum information available from the metric-space analysis, we computed $\bar{I}_{max}$, the average peak information available across all cases (see Methods). Across all metric-space analyses, $\bar{I}_{max}$ was 0.10 bits out of a possible 1.0 bit. As discussed below, this value suggests that additional information might be available in higher-level spike train features that cannot be captured by metric-space analyses. Additionally, since the proportion of rate and temporal cases did not differ significantly when computed from single- or multiunit data ($p>0.07$ in all cases; z-tests for proportions), we combined data from both types of recording in this as well as subsequent analyses. The similarity between the single- and multiunit datasets likely results from multiunit recordings in this paradigm only reflecting the activity of a single or a very small number of neurons, as discussed previously [14]. Finally, the results of the metric-space analysis were not sensitive to the number of behavioral groups used to classify the iterations of each song syllable. Although our primary analysis uses 2 behavioral groups (Fig. 1c, Fig. 2), as shown in Table 1 (Supplementary Information) we found a similar prevalence of rate and temporal cases when the trials were divided into three, five (Fig. 1d), or eight groups.

Our metric-space analysis therefore indicates that in most RA neurons, taking the fine temporal structure of spike trains into account provides better predictions of trial-by-trial variations in behavior than an analysis of spike rate alone (asterisks, Fig. 2d). Furthermore, at least when vocal outputs are grouped in three-dimensional acoustic space, spike timing can predict vocal acoustics significantly more frequently than would be expected from chance (cross, Fig. 2d). Although the latter result demonstrates that spike timing can carry significant information about vocal acoustics, it remains unclear whether spike timing can provide information about single acoustic parameters (beyond the 3D features). Answering this necessitates the direct method of calculating information, as described below.

*Direct method of calculating information.*

In the metric-space analysis, not all cases were classified as temporal. Further, when behavior was grouped by a single acoustic parameter rather than in multidimensional acoustic space, the number of temporal cases was not significantly larger than by chance (Fig 2d, green, yellow, and red plots). Thus it still remains unclear to what extent spike timing is important to this system overall, rather than in particular instances. Additionally, a drawback of metric-space analyses is that they assume that a particular model (metric) of neural activity is the correct description of neural encoding. As discussed more fully in Methods, metric-space approaches therefore provide only a lower bound on mutual information [23,25]. Put another way, metric-space analyses assume that the differences between spike trains can be fully represented by a particular set of parameters, which in our case include the temporal separation between nearest-neighbor spike times (Fig. 2a). However, if information is contained in higher order aspects of the spike trains that cannot be captured by these parameters (e. g. patterns that extend over multiple spikes), then metric-space analyses can significantly underestimate the true information contained in the neural code. We therefore estimated the amount of information that can be learned about the acoustic group by directly observing the spiking pattern at different temporal resolutions (Fig. 3a), without assuming a metric structure, similar to prior approaches in sensory systems [8,10]. We used the Nemenman-Shafee-Bialek (NSB) estimator to quantify the mutual information [26,27]. As described in Methods, this technique provides minimally biased information estimates, quantifies the uncertainty of the calculated information, and typically requires square-root-less data for estimation than many other direct estimation methods [26]. Nevertheless, the NSB technique requires significantly larger datasets than metric-space methods. We therefore directly computed mutual information using the subset (41/125) of cases where the recordings were long enough to gather sufficient data to be analyzed with this method.

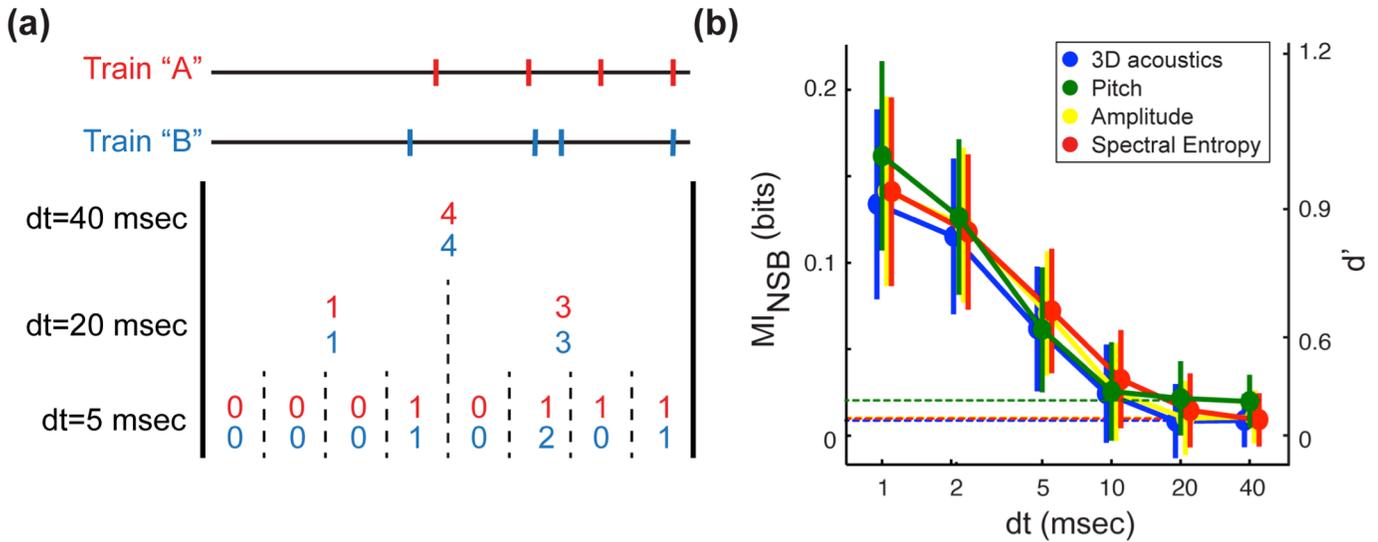

**Figure 3. Direct calculation of information reveals more information at finer temporal resolution. a,** The 40 msec-long spike train prior to each song syllable was converted into "words" with different time resolutions (*dt*), where the symbols within each word represent the number of spikes per bin. At *dt*=40 msec, two spike trains ("A" and "B") from our dataset are both represented by the same word ([4]). However, when *dt* decreases to 5 msec, the trains A and B were represented by different words ([0 0 0 1 0 1 1 1] and [0 0 0 1 0 2 0 1], respectively). We used the Nemenman-Shafee-Bialek (NSB) entropy estimation technique to directly compute the mutual information between the distribution of words and vocal acoustics at different temporal resolutions (see Methods). **b,** Mutual information ($MI_{NSB}$) increases as *dt* decreases. There is close to no information in the spike count, *dt*=40. Right-hand vertical axis shows information values converted into d' units. Error bars represent 1 SD of the information estimate. Here, the number of acoustic groups is 2 and the maximum possible information is therefore 1 bit. Dashed lines indicate mutual information at the 40 msec timescale and illustrate the mutual information expected at *dt*<40 if no information were present at faster timescales (i.e. from a rate code; see text).

We found that mutual information rose dramatically as temporal resolution increased. As shown in Figure 3b, when averaged across all 41 cases analyzed using the NSB technique, mutual information was relatively low when only spike counts were considered (i.e., for *dt*=40 msec). Across the four methods of grouping trials based on syllable acoustics, mutual information between spike counts and acoustic output ranged from 0.009-0.020 bits (with standard deviations of ~0.015), which is not significantly different from zero. If information about motor output were represented only in spike counts within the 40 msec premotor window, then mutual information at *dt*<40 would be equal to that found at *dt*=40 (dashed lines in Fig. 3b); note that this is true despite the increase in word length at smaller *dt* [8,10]. However, in all analyses mutual information increased as time bin size *dt* decreased and reached a maximum value at *dt*=1 msec, the smallest bin size (and thus greatest temporal resolution) we could reliably analyze. At 1 msec resolution, mutual information ranged from 0.134 -0.162 (with standard deviations of ~0.04) bits across the four analyses performed. These values of mutual information correspond to d' values near zero at *dt*=40 msec and to d' values between 0.9 and 1.0 at one-millisecond resolution (Fig. 3b, right-hand axis). These results indicate that far more information about upcoming vocal behavior is available at millisecond timescales and suggest that small differences in spike timing can significantly influence motor output. Therefore, although in some individual cases more information may be available from a rate code (empty bars, Fig. 2d), across the population of RA neurons much more information is present in millisecond-scale spike timing.

The results shown in Figure 3 demonstrate that millisecond-scale differences in spike timing within bursts can encode differences in behavior. To highlight these timing differences, we examined particular "words" (spike patterns) and considered how different timing patterns could predict vocal acoustics. Figure 4a and b each show 8 different words from a single neuron's response, color-coded according to the behavioral group in which each word appears most frequently. All words shown in Figure 4 contain the same number of spikes, and thus are identical at the time resolution of $dt$ = 40 msec (Fig. 3a). In the example shown in Figure 4a, a distinct set of spike timing patterns predicts the occurrence of low-pitched (group 1) or high-pitched (group 2) syllable renditions. In Figure 4b, behavioral groupings are performed in the three-dimensional acoustic space and similarly show that distinct spike timing patterns can predict vocal acoustics. In some cases, the timing patterns associated with behavioral groups share intuitive features. For example, the words associated with higher pitch in Figure 4a (blue boxes in grid) have shorter inter-spike-intervals than words associated with lower pitch (Fig. 4a, red boxes), suggesting that fine-grained interval differences drive pitch variation. However, in other cases (e.g., Fig. 4b) no such common features were apparent. Future studies incorporating realistic models of motor neuron and muscle dynamics are therefore required to understand how the precise timing patterns in RA can evoke differences in vocal behavior.

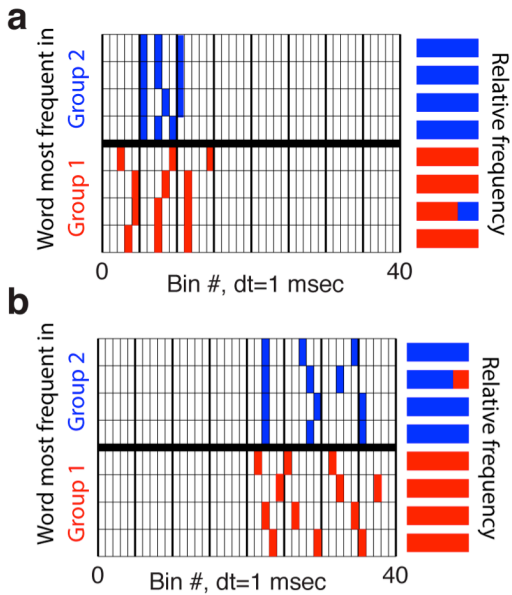

**Figure 4: Spike patterns within bursts predict vocal acoustics.** Each grid shows eight "words" at time resolution $dt$=1 msec (see Fig. 3a). Here we consider words with equal numbers of spikes (three). Rows represent different words, columns represent characters within a word, and boxes are filled when a spike is present. Words are color-coded according to which behavioral group they appear in most frequently, with words appearing more often in groups 1 and 2 shown in red and blue, respectively. Colored bars at right show the relative frequency with which each word appears in group 1 or 2, for example a solid red bar indicates a word that only occurs in behavioral group 1. Data in **a** are from the same neuron shown in Figure 1b, with behavioral groups determined by pitch (Fig. 1c). Data in **b** are from a different neuron with behavioral grouping in 3D acoustic space. Note that although this figure illustrates subsets of observed words, mutual information is always computed over the full distribution of all words.

*Comparing information estimates across analyses*

We compared the maximum information available from the metric-space analysis (see Methods), which is $\bar{I}_{max}$=0.10 bits, to the information available at the smallest $dt$=1 msec in the direct information calculation, $MI_{NSB}$=0.16 bits. Reassuringly, the peak information available from the direct method is of the same order of magnitude but somewhat larger than that computed independently in the metric-space analysis. This points at consistency between the methods and yet suggests that additional information may be present in higher order spike patterns that cannot be accounted for by a metric-space analysis, namely in temporal arrangements of three or more spikes. Similarly, a common technique in metric-space analysis is to estimate the "optimal time scale" of encoding as $1/q_{max}$ (although other authors suggest that such estimates may be highly imprecise[25]). In our dataset, the median value of $q_{max}$ was 0.3 msec$^{-1}$, suggesting that spike timing precision is important down to $1/q_{max} \sim 1$ msec, which is again in agreement with the direct estimation technique.

## Discussion:

We computed the mutual information between premotor neural activity and vocal behavior using two well-established computational techniques. A metric-space analysis demonstrated that spike timing provides a better prediction of vocal output than spike rate in a significant majority of cases (Fig. 2). A direct computation of mutual information, which was only possible in the subset of recordings that yielded relatively large datasets, revealed that the amount of information encoded by neural activity was maximal at a 1 msec timescale, while the average information available from a rate code was insignificant (Fig. 3). It also suggested that information in the spike trains may be encoded in higher order spike patterns.

Although previous studies have shown that bursts in RA projection neurons are aligned in time to the occurrence of particular song syllables [15,18], ours is the first demonstration that variations in spike timing within these bursts can predict trial-by-trial variations in vocal acoustics. These acoustic variations are thought to underlie vocal learning ability in songbirds. A number of studies have demonstrated that nucleus LMAN (the lateral magnocellular nucleus of the anterior nidopallium), the output nucleus of the AFP and an input to RA (Fig. 1a), both generates a significant fraction of vocal variability and is required for adaptive vocal plasticity in adult birds [28-30]. A significant question raised by our results therefore concerns the extent to which LMAN inputs can alter the timing of spikes in RA. Recent work has shown that spike timing patterns in LMAN neurons encode the time during song [31]. Future studies might address whether the observed patterns in LMAN spiking can also predict acoustic variations, and lesion or inactivation experiments could quantify changes in the distribution of firing patterns in RA after the removal of LMAN inputs [32].

Our results indicate that spike timing in cortical motor networks can carry significantly more information than spike rates. Equivalently, these findings suggest that limiting the analysis of motor activity to spike counts can lead to drastic underestimates of information. This contrast is illustrated by a comparison of the present analysis and our prior study examining correlations between premotor spike counts and the acoustics of song syllables [14]. In that earlier study, we found that spike rate predicted vocal output in ~24% of cases, a prevalence similar to the proportion of rate cases observed in the metric-space analysis and far smaller than the prevalence of temporal cases (Fig. 2). Similarly, direct computations of mutual information (Fig. 3) show that a purely rate-based analysis would detect only a small fraction of the information present in millisecond-scale timing. Therefore our central finding – that taking spike timing into account greatly increases the mutual information between neural activity and behavior – suggests that correlation and other rate-based approaches to motor encoding might in some cases fail to detect the influence of neural activity on behavior.

As shown in Figure 3, we found that spike timing at the 1 msec timescale provides an average of ~0.15 bits out of a possible 1.0 bit of information when discriminating between two behavioral groups. While this value is of course less then the maximum possible information, it is important to note that this quantity represents the average information available from a *single* neuron. A number of studies in sensory systems have demonstrated that ensembles of neurons can convey greater information than can be obtained from single neurons [33]. While our dataset did not include sufficient numbers of simultaneous recordings to address this issue, future analyses of ensemble recordings could test the limits of precise temporal encoding in the motor system.

Temporal encoding in the motor system could also provide a link between sensory processing and motor output. Prior studies have shown that different auditory stimuli can be discriminated based on spike timing in auditory responses [11,34,35], including those in area HVC, one of RA's upstream inputs [36]. Our results demonstrate that in songbirds, temporally precise encoding is present at the motor end of the sensorimotor loop. Propagating sensory-dependent changes in spike timing into motor circuits during behavior might therefore underlie online changes in motor output in response to sensory feedback [37,38] or serve as a substrate for long-term changes in motor output resulting from spike timing-dependent changes in synaptic strength [19,39,40].

While the existence of precise spike timing is strongly supported for a variety of sensory systems, a lingering question is how downstream neural networks could use the information that is present at such short timescales, and hence whether the animal's behavior could be affected by details of spike timing. Although theoretical studies have suggested how downstream neural circuits could decode timing-based spike patterns in sensory systems [41], the general question of whether the high spiking precision in sensing, if present, is an artifact of neuronal biophysics or a deliberate adaptation remains unsettled [42].

In motor systems, in contrast, spike timing differences could be "decoded" via the biomechanics of the motor plant, thereby transforming differences in spike timing into measureable differences in behavior. In a wide range of species [43-46], the amplitude of muscle contraction can be strongly modulated by spike timing differences in motor neurons (i.e., neurons that directly innervate the muscles) owing to strong nonlinearities in the transform between spiking input and force production in muscle fibers. Furthermore, biomechanical studies have shown that vocal muscles in birds have extraordinarily fast twitch kinetics and can reach peak force production in less than 4 msec after activation [20,21], suggesting that the motor effectors can transduce millisecond-scale differences in spike arrival into significant differences in acoustic output. Finally, *in vitro* and modeling studies have quantified the nonlinear properties the songbird vocal organ, demonstrating that small differences in control parameters can evoke dramatic and rapid transitions between oscillatory states, suggesting again that small differences in the timing of motor unit activation could dramatically affect the acoustics of the song [47,48].

Our results demonstrate that the temporal details of spike timing, down to 1 msec resolution, carry about ten times as much information about upcoming motor output compared to what is available from a rate code. This is in marked contrast to sensory coding [8,10], where the information from spike patterns at millisecond resolution is often about double that available from the rate alone. For this reason, the most striking result of our analysis might be that precise spike timing in at least some motor control systems appears to be even more important than in sensory systems. In summary, although future work in both sensory and motor dynamics is need to fully explicate how differences in spike timing are mapped into behavioral changes, our findings, in combination with previous results from sensory systems, represent the first evidence for the importance of millisecond-level spiking precision in shaping behavior throughout the sensorimotor loop.

48      Fee, M. S., Shraiman, B., Pesaran, B. & Mitra, P. P. The role of nonlinear dynamics of the syrinx in the vocalizations of a songbird. *Nature* **395**, 67-71, (1998).

**Methods:**

To measure the information about vocal output conveyed by motor cortical activity at different timescales, we recorded the songs of Bengalese finches while simultaneously collecting physiological data from neurons in RA. We then quantified the acoustics of individual song syllables and divided the iterations of each syllable into "behavioral groups" based on acoustic features such as pitch, amplitude, and spectral entropy. Mutual information was then computed using two complementary techniques. First, we used a metric-space analysis [1] to quantify how well the distance between pairs of spike trains can be used to classify syllable iterations into behavioral groups. Second, we used a direct calculation of mutual information [2-5] to produce a minimally-biased estimate of the information available at different timescales.

*Neural recordings*

Single-unit and multiunit recordings of RA neurons were collected from four adult (>140 days old) male Bengalese finches using techniques described previously [6]. All procedures were approved by the Emory University Institutional Animal Care and Use Committee. Briefly, an array of four or five high-impedance microelectrodes was implanted above RA nucleus. We advanced the electrodes through RA using a miniaturized microdrive to record extracellular voltage traces as birds produced undirected song (i.e. no female bird was present). We used a previously-described spike sorting algorithm [6] to classify individual recordings as single-unit or multiunit. In total, we collected 53 RA recordings (19 single-unit, 34 multiunit), which yielded 34 single-unit and 91 multiunit "cases", as defined below. Based on the spike waveforms and response properties of the recordings, all RA recordings were classified as putative projection neurons that send their axons to motor nuclei in the brainstem [6-8]. A subset of these recordings has been presented previously as part of a separate analysis [6].

*Acoustic analysis and premotor window*

We quantified the acoustics of each song syllable as described in detail previously [6]. Briefly, we quantified the fundamental frequency (pitch), amplitude, and spectral entropy at a particular time when spectral features were well-defined (Fig. 1b, red line) during each iteration of a song syllable. We selected these three acoustic features because they capture a large percentage of the acoustic variation in Bengalese finch song [6]. For each iteration of each syllable, we analyzed spikes within a temporal window prior to the time at which acoustic features were measured. The width of this window was selected to reflect the latency with which RA activity controls vocal acoustics. Although studies employing electrical stimulation have produced varying estimates of this latency [9,10], a single stimulation pulse within RA modulates vocal acoustics with a delay of 15-20 msec [11]. We therefore set the premotor window to begin 40 msec prior to the time when acoustic features were measured and to extend until the measurement time (Fig. 1b, red box). This window therefore includes RA's premotor latency [6,12] and allows for the possibility that different vocal parameters have different latencies.

*Determining behavioral groups*

While grouping spike trains is straightforward in many sensory studies, where different stimuli are considered distinct groups, we face the problem of continuous behavioral output in motor systems. We took two approaches to binning continuous motor output into discrete classes. First, we considered only a single acoustic parameter and divided the trials into equally sized groups using all of the data. For example, Figure 2a shows trials divided into two behavioral groups based on one parameter (pitch). In addition to pitch, separate analyses also used sound amplitude or spectral entropy to divide trials into groups. In the second approach (which we term "3D acoustics"; Fig. 2b), we used k-means clustering to divide trials into groups. Clustering was performed in the three-dimensional space defined by pitch, amplitude, and entropy, with raw values transformed into z-scores prior to clustering. Note that both approaches allow us to divide the dataset into an arbitrary number of groups (parameter *N*, see "Discrimination analysis" below). Our primary analysis divided trials into *N*=2 groups since a smaller *N* increases statistical power by increasing the number of data points in each group. However, alternate analyses using greater *N* yielded similar conclusions (see Results).

*Information calculation I: Metric-space analysis*

In previous studies, metric-space analysis has been used to probe how neurons encode sensory stimuli (for a review, see [13]). The fundamental idea underlying this approach is that spike trains from different groups (e.g. spikes evoked by different sensory stimuli) should be less similar to each other than spike trains from the same group (spikes evoked by the same sensory stimulus). In the present study, we adapt this technique for use in the vocal motor system to ask how neurons encode trial-by-trial variations in the acoustic structure of individual song syllables. To do so, we divide the iterations of a song syllable into "behavioral groups" based on variations in acoustic structure (Fig. 1c). We then construct a "classifier" to ask how accurately each spike train can be assigned to the correct behavioral group using a distance metric that quantifies the dissimilarity between pairs of spike trains [14]. As described in detail below, the classifier attempts to assign each trial to the correct behavioral group based on the distances between that trial's spike train and the spike trains drawn from each behavioral group. Crucially, the distance metric is parameterized by *q*, which reflects the importance of spike timing to the distance between two spike trains. This method therefore allows us to evaluate the contribution of spike timing to the performance of the classifier, and thus to the information contained in the spike train about the behavioral group.

*Calculating distances*

The distance metric used in this study, $D[q]$, is a normalized version of the distance metric $D^{spike}[q]$ originally introduced by Victor and Purpura [14,15]. The original metric is defined as the minimal cost of transforming one spike train into a second. There are three elementary steps, insertion (Fig. 2a, green circle) and deletion (Fig. 2a, green 'X') of a spike, which have a cost of 1, and shifting a spike (Fig. 2a, green arrows), which has a cost that is directly proportional to the amount of time the spike is moved. The proportionality constant, $q$, can take on values from 0 to infinity. When $q=0$, there is no cost for shifting spikes, and the distance between two spike trains is simply the absolute value of the difference between the number of spikes in each. For $q>0$, spike timings matter, and distances are smaller when spike times are similar between the two spike trains. The distance is normalized by dividing by the total number of spikes from both spike trains. The normalized version of the Victor and Purpura distance is more consistent with the assumption that spike trains with the same underlying rate should have smaller distances than spike trains with different rates [15]. Importantly, the time-scale parametric nature of $D[q]$ allows us to evaluate the contribution of spike timing to the amount of information transmitted by the neuron about the behavior.

*Classifier-based measurement of mutual information*

To determine the amount of systematic, group-dependent clustering, a decoding algorithm ("classifier") is used to classify the spike trains into predicted groups based on $D[q]$. The performance of the classifier in discriminating between behavioral groups is measured by calculating the mutual information between the actual group and predicted group.

The classifier assigns trials to a predicted group by minimizing the average distance to the group. Given a spike train $s$, we calculate the average distance from $s$ to the spike trains pertaining to a certain group $G_i$ by:

$$d(s, G_i) = \left[\langle (D[q](s, s'))^z \rangle_{s' \text{from group } G_i}\right]^{1/z} \tag{1}$$

If $s$ belongs to group $G_i$, we exclude the term $D[q](s,s)$ from the above equation. The trial is classified into the group $G_i$ that minimizes this average distance, and the resulting information is summarized into a confusion matrix $C(G_k^P, G_l)$ which indicates the number of times that a trial from group $G_l$ is assigned to group $G_k^P$. The parameter $z$ determines the geometry of the average, biasing the average to the shortest distances for negative values and emphasizing reducing the distance to outliers for positive values.

This procedure is performed for a range of $q$ values (0, 0.05, 0.1, 0.2, 0.3, 0.5, 1, 2, 5, 10, and 20 msec$^{-1}$) to produce a set of confusion matrices, which are normalized into probability matrices $P(G_k^p, G_l)$ by dividing by the total number of spike trains. Then the performance of the classification can be measured by computing the mutual information, $I$, between the actual group and predicted group.

$$I(G^P, G; q) = \sum_{k=1}^{N}\sum_{l=1}^{N} p(G_k^P, G_l; q) \log_2 \frac{p(G_k^P, G_l; q)}{p(G_k^P; q)p(G_l; q)} \tag{2}$$

The variable N in Equation 2 refers to the number of groups each dataset's trials were divided into. Except where otherwise indicated, we used $N$=2. To optimize the performance of the classifier, we maximized mutual information across different values of $z$ in the range of -8 to 8 for each value of $q$, as described previously [15].

$I_{count}(G^P, G)$ is the information when only spike counts are considered, that is when $q$=0, or $I(G^P, G; 0)$. $I_{max}(G^P, G)$ is the maximum value of $I(G^P, G; q)$, and the value of $q$ associated with $I_{max}(G^P, G)$ is $q_{max}$. If $I(G^P, G; q)$ plateaus, obtaining $I_{max}(G^P, G)$ at more than one value of $q$, $q_{max}$ is defined as the smallest of those values.

*Bias correction, "classifier"*

Because there is a component of the classification that is correct by chance, the estimate from Equation 2 can overestimate the true information. This bias can be computationally approximated and subtracted from the original estimate [14]. Concretely, we shuffle the spike trains across groups and then perform the analysis 1000 times and calculate the average information across these random reassignments. This value is an estimate of the bias and is subtracted from the original estimate. After subtraction, only values above the 95th percentile of the null distribution of *I* values are considered significant and negative values are set equal to zero.

*Rate cases versus temporal cases*

We define a "case" as one neural recording (single- or multi- unit) that meets an average firing threshold of 1 spike in the 40ms premotor window before one syllable. We limited our analysis to cases for which at least 50 trials were available. After performing the above analyses on each case, we categorized the cases into "rate cases" and "temporal cases". Rate cases are when the maximum amount of information occurs for $q=0$. For rate cases, $I_{count} = I_{max}$, indicating that the best discrimination occurs when only spike counts are considered. For cases where $q_{max}>0$, the fine temporal structure of the spike train also contributes to discrimination, which we define as a temporal case.

      To determine whether the proportion of temporal cases, $\hat{p}_T$, is significantly greater than chance, we constructed synthetic datasets in which we randomized spike times for each trial in each case ("Poisson test"). These randomized spike trains had the same number of spikes as our original data. We then performed metric-space analysis in the same manner as before and calculated the proportion of

temporal cases across all cases. After generating one thousand of these synthetic datasets, we found the distribution of $p_T$ under the null hypothesis that spike timings do not encode motor output and asked whether our observed $\hat{p}_T$ was greater than the 95th percentile of this distribution. Additionally, we performed one-sided z-tests for proportions to ask whether the proportion of temporal cases exceeded the proportion of rate cases.

*Information calculation II: Direct method*

In addition to the metric-space analysis described above, we also directly calculated the mutual information between song acoustics and neural activity [5]. Whereas metric-space analysis makes strong assumptions about the structure of the neural code, the direct approach is model-independent [5,16]. Specifically, spike train distance metrics assume that spike trains that have spike timings closer to each other are linearly more similar than spike trains whose timings are more different. As with all assumptions, the methods gain extra statistical power if they are satisfied, but they may fail if the assumptions do not hold. The direct method simply considers distinct patterns of spikes at each timescale, without assigning importance to specific differences. Crucially, direct methods allow us to estimate the true mutual information, whereas the mutual information computed from a metric-space analysis represents only a lower bound on this quantity [17]. However, because the direct method is a model-independent approach that does not make strong assumptions about the neural code, it requires larger datasets to achieve statistical power.

To determine whether there is information about acoustics in the precise timing of spikes, we compared the information between neural activity and behavioral group following discretization of the spike trains at different time resolutions. For a time bin of size $dt$, each $T$ = 40 msec-long spike train was transformed into a "word" with $40/dt$ symbols where different symbols represent the number of spikes per bin. The mutual information is simply the difference between the entropy of the total distribution of words $H_{T,dt}[R]$ and the average entropy of the words given the behavioral group $H_{T,dt}[R|G]$:

$$I_{direct\ T,dt}[R;G] = H_{T,dt}[R] - \langle H_{T,dt}[R|G]\rangle_G \tag{4}$$

$I_{direct}$ could be quantified exactly if the true probability distributions $p(R)$, $p(R,G)$ and $p(R|G)$ were known:

$$I_{direct} = -\sum_R p(R) \log_2 p(R) - (-\sum_R \sum_G p(R,G) \log_2 p(R|G)). \tag{5}$$

However, estimating these distributions from finite datasets introduces a systematic error ("limited sample bias" or "undersampling bias") that must be corrected [18]. There are several methods to correct for this bias, but most assume that there is enough data to be in the asymptotic sampling regime, where each typical response has been sampled multiple times. As we increase the time resolution of the binning of the spike train, the number of possible neural responses increases exponentially, and we quickly enter the severely undersampled regime where not every "word" is seen many times, and, in fact, only a few words happen more than once (which we term "coincidence" in the data). We therefore employed the Nemenman-Shafee-Bialek (NSB) entropy estimation technique [2,4], which can produce unbiased estimates of the entropies in Equation 4 even for very undersampled datasets.

The NSB technique uses a Bayesian approach to estimate entropy. However, instead of using a classical prior, for which all values of the probability of spiking are equally likely, NSB starts with the a priori hypothesis that all values of the entropy are equally likely. This approach has been shown to reliably estimate entropy in the severely undersampled regime (where the number of trials per group is much less than the cardinality of the response distribution) provided that the number of coincidences in that data is significantly greater than one. This typically happens when the number of samples is only about a square root of what would be required to be in the well-sampled regime [2,3].

This method often results in unbiased estimates of the entropy, along with the posterior standard deviation of the estimate, which can be used as an error bar on the estimate [3]. On the other hand, we know that no method can be universally unbiased for every underlying probability distribution in the severely undersampled, square-root, regime [19]. Thus there are many underlying distributions of spike trains for which NSB would be biased. Correspondingly, the absence of bias cannot be assumed and must instead be verified for every estimate, which we do as described below.

We restricted our analysis to cases in which the number of trials was large enough (>200) so that the number of coincidences would likely be significantly greater than 1. Of our 125 datasets, 41 passed this size criterion. We emphasize that no additional selection beyond the length of recording was done. Since recording length is unrelated to the neural dynamics, we expect that this selection did not bias our estimates in any way. The NSB analysis was performed using $N$=2 behavioral groups, since increasing the number of groups greatly decreased the number of coincidences and increased the uncertainty of the entropy estimates (not shown). Because NSB entropy estimation assumes that the words are independent samples, we have to check that temporal correlations in the data are low. To do this, we used NSB to calculate the entropy of four different halves of each dataset: the first half of all trials, the second half, and the two sets of every other trial, where the second set is offset from the first set by one trial. If the difference in mean entropy between the first half and second half data were comparable to the difference between the two latter sets, then the effects of temporal correlations are low. This would mean that the correlations are unlikely to affect entropy estimation, and the information at high spiking precision that we observe cannot just be attributed to modulation occurring on a longer time scale.

To make sure that the NSB estimator is unbiased for our data, we estimated each conditional and unconditional entropy from all available $N$ samples, and then from $\alpha N, \alpha < 1$, samples. Twenty-five random subsamples of size $\alpha N$ were taken and then averaged to produce $\langle S(\alpha) \rangle$. We plotted $\langle S(\alpha) \rangle$ vs. $1/\alpha$ and checked whether all estimates for $1/\alpha \to 1$ agreed among themselves within error bars, indicating no empirical sample-size dependent bias [5,20]. In most cases, no sample size-dependent drift in the entropy estimates was observed, and hence the estimates from full data were treated as unbiased. In those cases where bias was visible, it could often be traced to the rank-ordered distribution of words not matching the expectations of the NSB algorithm. Specifically, some of the most common words occurred much more often than expected from the statistics of the rest of the words. Since NSB uses frequencies of common, well-sampled words to extrapolate to undersampled words, such uncommonly frequent outliers can bias entropy estimation [4]. To alleviate the problem, we followed [20] and partitioned the response distribution in a way such that the most common word was separated from the rest when it was too frequent (>2% of all words). We use $p_1$ to denote the frequency of the most common word and $p_2$ to denote the frequency of all other words. We then used the additivity of entropy,

$$S = p_1 S_1 + p_2 S_2 + S(p_1, p_2) \tag{6}$$

to compute the total entropy by first estimating the entropy of the choice between the most common word and all others, $S(p_1, p_2)$, and the entropy of most of the data, $S_2$, independently using the NSB method (the entropy of the single most common word, $S_1$, is zero). The error bars were computed by summing the individual error bars in Eq. (6) in quadratures. As verified by the subsampling procedure explained above, entropies of all but 5 cases were unbiased once isolation of the most common word was performed.

We then averaged the mutual information between the spike train and the acoustic group over all cases, weighing contribution of each case by the inverse of its respective posterior variance. The variance of the mean was similarly estimated. Since remaining biased cases were so few, and they typically had few coincidences and hence large error bars, we expect that these biased cases did not contribute significantly to bias the average mutual information.

*Peak information from metric-space method:*
As discussed above, the metric-space and direct methods of computing mutual information differ in their underlying assumptions about the statistical structure of the neural code, and the metric-space method can only produce lower bounds on the signal-response mutual information. Therefore, comparing the values of information computed by the two methods is prone to various problems of interpretation. It is nevertheless instructive to ask whether the direct method estimates greater mutual information than the metric-space analysis, and thus if patterns of multiple spikes carry additional information beyond that in spike pairs, which is discoverable by the metric-space method. To answer this, we calculated the peak metric space information $\bar{I}_{max}$, which is the mean of $I_{max}$ across all cases. This is the upper bound on the information detectable through the metric-space method, as the information is maximized for each case independently, rather than finding a single optimal *q* for all cases.

**Acknowledgements:**

We thank Michael Long and Robert Liu for helpful discussions and Harshila Ballal for animal care. This work was supported by US National Institutes of Health grants R90DA033462, P30NS069250, R01NS084844, and R01DC006636, National Science Foundation grant IOS-1208126, and McDonnel Foundation grant 220020321.

**Author contributions:**

C.T., I.N., and S.J.S. designed the study and wrote the paper. S.J.S. and D.C. collected the neural data, C.T., K.S., I.N., and S.J.S. analyzed the data.

**SUPPLEMENTARY INFORMATION**

**Table 1.** Effect of dividing trials for each case into a different number of behavioral groups. Numbers in the table are the percentages of total cases that are rate cases and temporal cases, respectively. Asterisks indicate instances where the proportion of temporal cases is significantly greater than the proportion of rate cases.

| N | 3d acoustics<br>R, T in % | Pitch<br>R, T in % | Amplitude<br>R, T in % | Spectral Entropy<br>R, T in % |
|---|---|---|---|---|
| 2 | 17.6, 53.6 | 16.8, 56.0* | 13.6, 61.6* | 16.8, 56.8* |
| 3 | 26.4, 53.6* | 35.2, 45.6 | 20.8, 48.0* | 20.0, 57.6* |
| 5 | 20.0, 54.4* | 26.4, 53.6* | 24.0, 50.4* | 31.2, 59.2* |
| 8 | 23.2, 54.4* | 24.8, 49.6* | 18.4, 61.6* | 20.0, 59.2* |